\documentclass[aps,prl,twocolumn,amsmath,amssymb,showpacs,floatfix]{revtex4}
\usepackage{graphicx}% Include figure files
\usepackage{dcolumn}% Align table columns on decimal point
\usepackage{bm}% bold math

\begin{document}

\title{Radiative Lifetime of Excitons in Carbon Nanotubes}

\author{Vasili Perebeinos$^*$, J. Tersoff, and Phaedon Avouris}
\affiliation{IBM Research Division, T. J. Watson Research Center,
Yorktown Heights, New York 10598}

\date{\today}

\begin{abstract}
We calculate the radiative lifetime and energy bandstructure
of excitons in semiconducting carbon nanotubes,
within a tight-binding approach.
In the limit of rapid interband thermalization, the radiative decay rate
is maximized at intermediate temperatures, decreasing at low temperature
because the lowest-energy excitons are optically forbidden.
The intrinsic phonons cannot scatter excitons between optically active
and forbidden bands,
so sample-dependent extrinsic effects that break the symmetries
can play a central role.
We calculate the diameter-dependent energy splittings
between singlet and triplet excitons of different symmetries,
and the resulting dependence of radiative lifetime
on temperature and tube diameter.
\end{abstract}

 % chosen 12Apr2005
\maketitle

The potential usefulness of carbon nanotubes as an optical material
depends sensitively on the luminescence efficiency,
which in turn depends on the radiative lifetime.
Yet while the optical properties of nanotubes have been studied intensely
\cite{Li,Connell,Bachilo,Hagen,Lebedkin,Lefebvre,Ando,%
Pedersen,Kane,Louie,Chang,Perebeinos1,Perebeinos4},
little is known about the radiative lifetime,
and its dependence on temperature and tube diameter.
It is by now well recognized that optical absorption and emission
in nanotubes are dominated by excitons \cite{Louie,Perebeinos1,Perebeinos4},
posing an additional challenge.

Here we calculate the exciton bandstructure and
radiative lifetime in carbon nanotubes.
When interband thermalization is efficient, we find that the
lifetime has a minimum at {\it intermediate} temperatures,
with optical recombination becoming slower both at very
low and very high temperature.
This unusual behavior occurs because the
lowest-energy excitons are optically forbidden by symmetry
\cite{Perebeinos1,Zhao}.
We report the energy differences between singlet and triplet excitons
of different symmetries, for a range of nanotube diameters.
We also present the resulting temperature dependences
of the radiative lifetime.
Both the energies and temperature dependences are well described
by simple scaling formulas.
The optically allowed band has a non-analytic dispersion
due to exchange, while the other bands have a mass enhancement
of $\sim$50\% over free electron-hole pairs.

We find that symmetry plays a special role in this system.
For photoluminescence of an ideal nanotube
(and neglecting spin-orbit coupling),
phonons cannot scatter excitons into the optically forbidden
low-energy states, so optical emission is very efficient.
However, extrinsic effects that break the ideal symmetry
can qualitatively change the behavior, especially at low temperature.
Such effects can also lead to multiple peaks in emission
spectra with characteristic temperature dependence,
perhaps explaining experimental observations \cite{linewidth,Hagen2}.

The quantum efficiency $\eta$ of light emission corresponds to the
fraction of excitons decaying by radiative channels:
$\eta=\tau_{nr}/(\tau_r+\tau_{nr})$, where $\tau_r$ and
$\tau_{nr}$ are the radiative and nonradiative lifetimes. The
total decay rate (i.e.\  inverse lifetime) is $\tau_{tot}^{-1} =
\tau_r^{-1} + \tau_{nr}^{-1}$. It has been measured
\cite{Ma,Ostojic,Reich,Wang,Vardeny}, and environmental effects
have been recently published \cite{Tobiasnew}. But only one
measurement of radiative lifetime has been reported~\cite{Wang}.

A photoexcited state has total momentum $q=q$(photon),
hereafter approximated as $q=0$.  It is rapidly scattered by
phonons into a different state with $q \ne 0$ \cite{Perebeinos4},
and cannot decay directly.
Therefore what is relevant is not the lifetime of a specific
excitonic state, but rather the lifetime of an exciton as it
scatters from one state to another.
If thermalization is inefficient, then the lifetime
depends on the details of
the excitation process and subsequent scattering pathways.

We therefore focus first on the case
where the exciton thermalizes on a timescale
much shorter than the decay time \cite{Ma,Ostojic,Reich,Wang,Vardeny}.
Then at low excitation density there are well-defined
(thermally averaged) radiative and non-radiative lifetimes.
The radiative decay rate (i.e.\ inverse radiative lifetime) is
\cite{Feldmann}
\begin{eqnarray}
\tau_r^{-1}=\frac{1}{Z(T)}\sum_{\nu}\tau_{\nu}^{-1}
\exp\left(  \frac{-E_{\nu}}{k_{B}T}\right),
\label{eq1}
\end{eqnarray}
and a similar formula applies for nonradiative decay. The index
$\nu=(q,j,S,\sigma)$ may include summation over exciton wavevector
$q$, angular momentum $j$ (i.e.\  discrete wavevector around the
tube), spin index $\sigma$ (labelling singlet and triplet states),
and exciton state $S$ (labelling the bound and continuum states
for a given $q$ and $\sigma$); or it may be restricted to those
states accessible by symmetry-allowed scattering mechanisms, as
discussed below. The partition function is
$Z(T)=\sum_{\nu}\exp(-E_{\nu}/k_{B}T)$, summed over the same range
of $\nu$. The radiative decay rate for each excitonic state is
\cite{Stern}
\begin{eqnarray}
\tau_{\nu}^{-1}=\frac{ne^{2}E_{\nu}^{2}f_{\nu}}
{2\pi\varepsilon_{0}m_{e}\hbar^{2}c^{3}}
\label{eq2}
\end{eqnarray}
where $n$ is the refractive index (hereafter $n=1.4$). We calculate
the exciton energies $E_{\nu}$ and the corresponding oscillator
strengths $f_{\nu}$ using a tight-binding Hamiltonian
\cite{Saito} with $t=3.0$ eV, and solving the Bethe-Salpeter Equation (BSE)
\cite{Rohlfing}, as in Refs.~\cite{Perebeinos1,Perebeinos4},
taking the surrounding medium to have dielectric constant $\varepsilon=2$.

In an idealized picture, one exciton can capture
much of the spectral weight of the entire nanotubes \cite{Hanamura}.
But in reality, phonon scattering limits the
exciton coherence to a finite length scale,
and spreads the oscillator strength of the
exciton over a finite linewidth \cite{Feldmann}.
As long as the linewidth is substantially less than $k_B T$
\cite{linewidth},
we obtain good accuracy from Eq.~(\ref{eq1})
% length: omit:
without including this linewidth explicitly.

\begin{figure}
\includegraphics[height=2.8in]{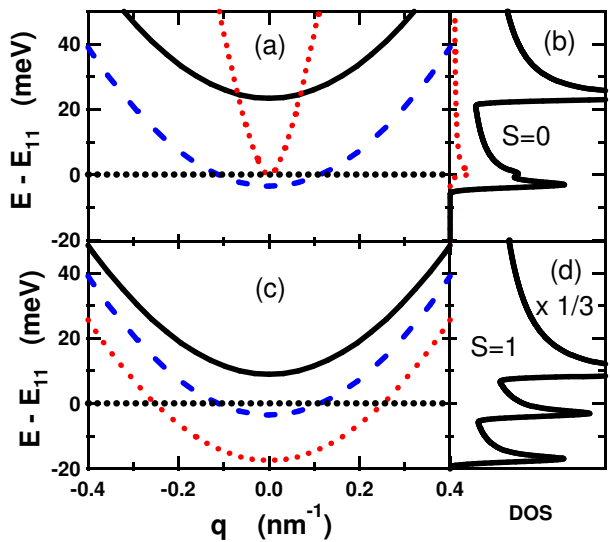}
\caption{\label{fig1} Four lowest-energy exciton bands of a (19,0)
nanotube ($d=1.5$ nm) embedded in medium having $\varepsilon=2$
\protect{\cite{Perebeinos1}}, for (a) singlet and (c) triplet spin
states. The black solid curves are doubly degenerate exciton bands
having nonzero angular momentum. The blue dashed curves are even
parity (dipole forbidden) exciton bands having zero angular
momentum. The red dotted curves are the odd parity (dipole
allowed) exciton bands having zero angular momentum. The density
of states for (b) singlet and (d) triplet states are shown by the
black curves. The red dotted curve in (b) shows the partial
density of states for dipole allowed singlet exciton. The change
in scale of 1/3 in (d) is for triplet degeneracy.}
\end{figure}

For a single 1D exciton band with parabolic dispersion,
the temperature dependence of the radiative lifetime [Eq.~(\ref{eq1})]
is $\tau_r \propto T^{1/2}$ \cite{Citrin}.
However, as shown in Fig.~\ref{fig1},
here there are several low-lying bands,
with the lowest being optically forbidden \cite{Perebeinos1,Zhao}.
Thus it is important to understand the exciton energies
and symmetries in some detail.

The conduction and valence bands in carbon nanotubes are doubly
degenerate. For free electron-hole pairs, this gives four distinct
but degenerate pair excitations.  In excitons, the Coulomb
interaction partially lifts the fourfold degeneracy. The
dispersions of the four lowest-energy singlet excitons are shown in
Fig.~\ref{fig1}a. The lowest-energy exciton at $q=0$ has zero angular
momentum and even parity, so it is dipole forbidden \cite{Perebeinos1}.
The second exciton also has zero angular momentum, but
with odd parity, and it gives rise to the $E_{11}$ optical
transition. The next two excitons at $q=0$ lie above the $E_{11}$
transition energy.  They are degenerate, with positive and negative
angular momentum, and are therefore optically forbidden;
but they can be observed in phonon-assisted absorption \cite{Perebeinos4}.

The optically forbidden excitons
all have similar effective masses,
$m^* \sim 1.5 \left( m_e + m_h \right)$,
where $m_e = m_h$ is the effective mass
of an electron or hole.
% in a tube of that diameter.
The 50\% mass enhancement is roughly independent of tube diameter.

In contrast, the dispersion of the optically active exciton is
extremely steep and non-parabolic, going as $q^2\ln\vert q\vert$.
A similar logarithmic contribution to the electron self energy was
found in 2D graphene \cite{Kane2}. This anomalous dispersion
occurs because of the exchange energy, which for the other bands
is either zero by symmetry, or is a smooth function of $q$.

\begin{figure}
\includegraphics[height=3.4in]{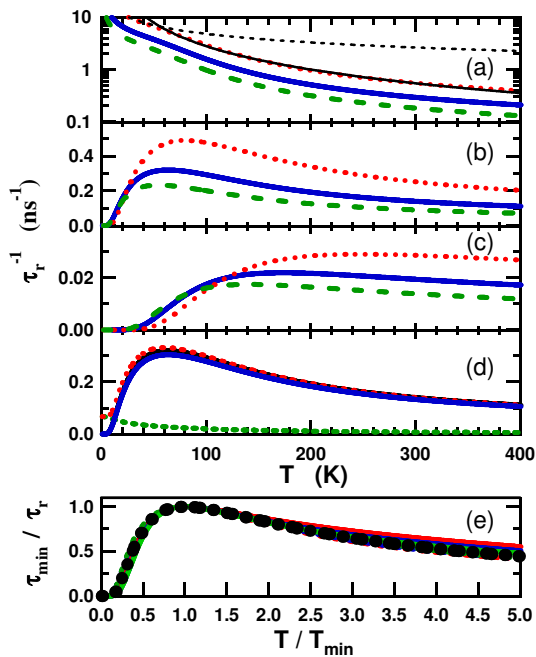}
\caption{\label{fig2} Exciton radiative decay rate $\tau_r^{-1}$
[Eq.~(\ref{eq1})] as a function of temperature, in a medium having
$\varepsilon=2$ \cite{Perebeinos1}. Curves in (a-c) are for
zig-zag tubes with diameters 1.0 nm (dotted red), 1.5 nm (solid
blue), and 2.0 nm (dashed green). Fine dotted black line in (a)
shows fit of 1.0 nm curve to $T^{-0.55}$ at low $T$ and $T^{-3/2}$
at high $T$. In (a) only singlet states of odd parity are
included; (b) assumes thermalization between singlet states of
even and odd parity; and (c) assumes thermalization between both
singlet and triplet states of both parities. Effect of symmetry
mixing is illustrated in (d) for 1.5 nm tube as in (b), but
including a 5\% mixing of spectral weight into the even-parity
band. Solid blue curve shows decay rate from main $E_{11}$ peak,
which is only very slightly reduced relative to (b). Dotted green
line is contribution from nominally forbidden exciton 3.5 meV
lower in energy.  Dashed red line is total radiative decay rate.
Temperature dependence of dotted green curve depends on
$q$-dependence assumed for mixing. (e) Scaled radiative lifetimes
for all six curves from (b) and (c). The empirical scaling formula
Eq.~(\protect{\ref{eq3}}) is shown by a dotted black-on-white
curve. }
\end{figure}

In photoluminescence experiments, excitons are optically pumped
into a singlet spin state of odd parity. For an ideal nanotube
(and neglecting spin-orbit coupling), we find that phonons cannot
scatter an exciton between states of different parity or different
spin. Therefore the exciton will thermalize within the odd-parity
spin-singlet bands (solid black and dotted red in
Fig.~\ref{fig1}a), and the sums in Eq.~(\ref{eq1}) and $Z(T)$ need
to be restricted accordingly. The resulting behavior is shown in
Fig.~\ref{fig2}a. At low temperature, the behavior is well
approximated as $\tau_{\rm r} \propto T^{0.55}$, slightly
different than $T^{1/2}$ dependence \cite{Citrin} due to the
nonparabolic dispersion. At higher temperature, the behavior is
quite different, reflecting the contribution of the higher-energy
bands to $Z(T)$; it is well approximated by $\tau_{\rm r} \propto
T^{3/2}$.

This ideal picture, while appealing in its simplicity, may not be
directly applicable to experiment. There are many effects that
tend to break the symmetries of the perfect nanotube: coupling to
the environment, structural distortions, finite size and
associated end effects, and the presence of defects or impurities.
These effects allow scattering and thermalization between bands of
different parity and/or different spin. This tends to suppress
luminescence at low temperature, because excitons thermalize to a
low-energy band that is optically forbidden. At the same time,
luminescence is suppressed at high temperature because a larger
range of $q$ becomes thermally accessible. These effects enter via
$Z(T)$ in Eq.~(\ref{eq1}).

Fig.~\ref{fig2}b shows the case of scattering and thermalization
between different parity, but not between different spin states;
and Fig.~\ref{fig2}c shows the case of thermalization among states
of all parities and spins. We find that the temperature dependence
of $\tau_r$ is very similar in all cases where thermalization
includes a low-energy forbidden band, except for an overall
scaling of $\tau$ and $T$. This is shown in Fig.~\ref{fig2}e,
where all the curves in Figs.~\ref{fig2}b-c collapse nearly to a
single scaled curve. A simple interpolation formula can describe
this scaled temperature dependence:
\begin{eqnarray}
\frac{\tau_r}{\tau_{\rm min}}=\frac{T}{T_{\rm
min}}\exp{\left(\frac{T_{\rm min}-T}{T}\right).}
 \label{eq3}
\end{eqnarray}

\begin{figure}
\includegraphics[height=2.7in]{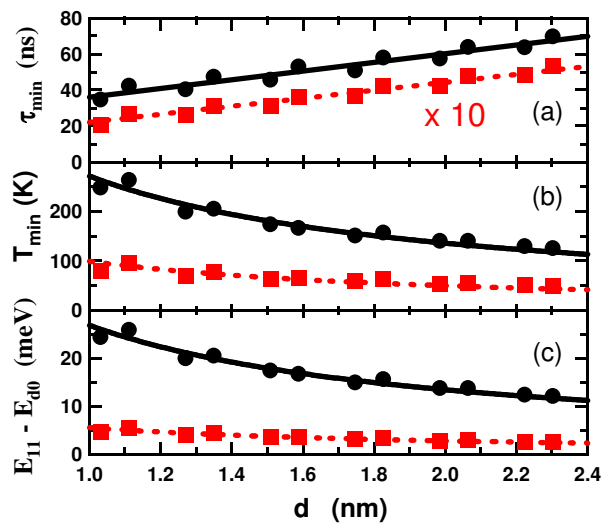}
\caption{\label{fig3} Diameter dependence in zig-zag
semiconducting nanotubes of (a) the minimum radiative lifetime,
(b) the corresponding temperature, and (c) the energy splitting
between the optically active exciton and the lowest energy
exciton, along with best fits to
Eq.~(\protect{\ref{eq45}}$-$\protect{\ref{eq6}}). Black circles
are the results, when both triplet and singlet states are
included; red squares are with only singlet excitons included.
Thus the black circles in (c) correspond to the singlet-triplet
splitting of the dipole-allowed states at $q=0$. Solid curves are
interpolation formulas, Eq.~(\protect{\ref{eq45}}) and
Eq.~(\protect{\ref{eq6}}). Note the $\times$10 multiplication
factor for the red squares in (a).}
\end{figure}

The diameter dependence of $\tau_{\rm min}$ and $T_{\rm min}$
are shown in Fig.~\ref{fig3}a and \ref{fig3}b respectively,
along with the best fits to the simple empirical relations:
\begin{eqnarray}
\tau_{\rm min}  \approx  \tau_0+\alpha d, ~~~~~~~
T_{\rm min}  \approx  \frac{\beta}{d}. \label{eq45}
\end{eqnarray}
For small-diameter (large-bandgap) tubes,
the maximum radiative decay rate $\tau^{-1}_{\rm min}$ is higher,
and this maximum rate occurs closer to room temperature.

$T_{\rm min}$ is approximately proportional to the energy difference
between the optically active exciton ($E_{11}$),
and the lowest-energy exciton that participates in thermalization
($E_{d0}$).
This difference $E_{11} - E_{d0}$ is shown in Fig.~\ref{fig3}c.
The singlet-triplet exchange
splitting of the dipole-allowed exciton, as well as the even-odd
parity splitting of the singlet exciton, are both inversely
proportional to the tube diameter:
\begin{eqnarray}
E_{11}-E_{d0}\approx\frac{\gamma}{d}. \label{eq6}
\end{eqnarray}
The singlet-triplet splitting is expected to scale with the size
of the exciton, which is proportional to
the tube diameter \cite{Perebeinos1}.
The splitting between the even and odd parity
excitons is a factor of five smaller than the singlet-triplet
exchange splitting.

In Fig.~\ref{fig3}, $\tau_{\rm min}$, $T_{\rm min}$, and
$E_{11}-E_{d0}$ all exhibit alternations in the diameter dependence,
due to the systematic difference between tubes having
mod(n-m,3)=1 or 2.  While only zigzag tubes are shown here,
we find very similar diameter dependence for other chiralities.

Equations (\ref{eq3}$-$\ref{eq45}) provide an estimate of the
temperature dependent radiative lifetime in this tube diameter
range, whenever thermalization with optically forbidden symmetries
occurs. The best-fit values of the three parameters are:
$\tau_0=12$ ns, $\alpha=24$ ns/nm, and $\beta=272$ K-nm, if both
singlet and triplet excitons contribute to $Z(T)$; and $\tau_0=0$
ns, $\alpha=2.2$ ns/nm, and $\beta=99$ K-nm if only singlet
excitons contribute. The singlet-triplet exchange splitting and
the even-odd parity splitting at $q=0$ are given by
Eq.~(\ref{eq6}), with $\gamma=27$ meV-nm and $\gamma=5.5$ meV-nm
respectively.

At sufficiently low temperature, additional complications arise.
Thermalization by acoustic phonons becomes less effective
\cite{Perebeinos3}, so it is less likely that this equilibrium
picture will apply. Also, even weak disorder from environmental
interactions would lead to localization of band-edge excitons.
Because localized excitons that are close in energy can be well
separated spatially, they typically are not in thermal equilibrium
at low $T$. This will give a finite linewidth determined by the
strength of the disorder, even at $T=0$.

Interband scattering necessarily implies
some mixing of the bands;
so in the cases of inter-symmetry scattering (Fig.~\ref{fig2}b-c),
the optically forbidden bands are no longer strictly forbidden.
Rather, there is some small but finite radiative decay rate
$\tau_\nu^{-1}$ associated with every participating band.
For thermalization at sufficiently low $T$,
only the lowest-energy of these bands
will have significant occupancy.  Then the emission from
this ``nearly-forbidden'' band, however small, will dominate
over thermal excitation into the optically allowed band.

In this case,
the emission spectrum (not shown) has two peaks,
one from the optically allowed band, and one at lower energy
from the nominally forbidden band.
The latter may be negligibly weak at high temperature,
but becomes dominant at sufficiently low temperature
(as long as thermalization occurs before decay).
We illustrate this in Fig.~\ref{fig2}d, for one particular
model of the mixing.

Thus the behavior is sensitive to deviations from ideality, which
will in general vary among different experiments. Single-tube
spectroscopy at low temperature provides a powerful diagnostic of
such extrinsic effects.  Any distinct emission peak at lower
energy than the main peak implies some symmetry breaking.  If the
peaks do not correspond to the few distinct bands of
Fig.~\ref{fig1}, they are likely to correspond to excitons that
are localized (e.g.\  at defects or deformations). A finite peak
width as $T \rightarrow 0$ suggests disorder (along with the
expected breakdown of thermalization). Recent measurements are
suggestive in showing multiple peaks, with spectral weight
shifting from high- to low-energy peaks with decreasing
temperature \cite{linewidth,Hagen2}.

\newpage

\end{document}